\documentstyle[12pt]{article}

\def\lapproxeq{\lower .7ex\hbox{$\;\stackrel{\textstyle <}{\sim}\;$}}
\def\gapproxeq{\lower .7ex\hbox{$\;\stackrel{\textstyle >}{\sim}\;$}}

\begin{document}

\titlepage

\begin{center}
\vspace*{1cm}
{\large{\bf  Diffractive vector meson production at large momentum transfer}}
\end{center}
\vspace*{.5cm}
\begin{center}
J.R. Forshaw \\
Rutherford Appleton Laboratory, \\ Chilton, Didcot OX11 0QX, England. \\
M.G. Ryskin\footnote{Work supported in part by the Russian Fund of
Fundamental Research (93-02-03145) and by the Volkswagen-Stiftung.}\\
St. Petersburg Nuclear Physics Institute\\
188350, Gatchina, St. Petersburg, Russia.\\
\end{center}

\vspace*{0.5cm}

\begin{abstract}
The diffractive process $\gamma p\to V + X$ (where $V$ is a
vector meson and $X$ results from the dissociation of the proton) is
studied. In particular, we consider the region of large momentum transfer
(i.e. $|t|\gg\Lambda_{QCD}^2$) and large centre-of-mass (CM) energy, $s$.
The asymptotic $(s \to \infty, s/|t|\gg 1$) behaviour is derived from the BFKL
equation and compared to that which is obtained in the Born approximation
(two-gluon exchange). We also calculate the corrections to the Born graphs by
iterating the BFKL kernel numerically. Improved convergence of the BFKL series
is found by summing the logarithms which occur when an exchanged gluon goes
nearly on shell. Importantly, we find evidence that the asymptotic solution to
the BFKL equation is inappropriate over most of the HERA range and we provide
more realistic predictions for the cross section. The predicted cross section
is not too small and can be measured at HERA, up to momentum transfers
$|t| \sim 10$GeV$^2$.
\end{abstract}

\newpage

\section{Introduction}
We consider the diffractive production of vector mesons in $\gamma p$
interactions, where the momentum transfer $-t \gg \Lambda_{QCD}^2$ and the
photon can be either real or virtual. The vector
meson ($V$) is produced quasi-elastically while the proton dissociates into a
diffracted mass, $M$ (this contribution dominates over the elastic channel at
large $t$). The scattering is mediated by the exchange of a colour
singlet object, i.e. the perturbative (`hard') pomeron \cite{fkl,lip}. The
largeness of the momentum transfer is expected to guarantee the dominance of
short distance physics and hence the use of perturbative QCD. As we shall
show, the reaction can be measured up to $-t \sim 10$ GeV$^2$ at
HERA and so we have the possibility to study the hard pomeron in detail.
In particular, there is the prospect to examine the QCD pomeron
trajectory away from $t=0$. Elastic (i.e. without proton dissociation)
vector meson production by photons
at low momentum transfers has been studied in refs.\cite{dl,cud,ry,bf}.

We start by considering the amplitude corresponding to Feynman graphs like
those of figs.1(a,b). The contributions from the
more complicated graphs, such as those in figs.1(c,d), will be discussed in
section 4. Thus we can factorise the cross section into a product of the usual
parton distribution functions and a quasi-elastic `hard scattering' cross
section:
\begin{equation}
\frac{d\sigma^T(\gamma^*p \to V+X)}{dt dx'} = \left
( \frac{81}{16}G(x',t)+\sum_f(q(x',t)\,+\,\bar{q}(x',t))\right )
\frac{d\sigma^T(\gamma^* q \to V q)}{dt}
\end{equation}
and
\begin{equation}
\frac{d\sigma^T(\gamma^* q \to V q)}{dt} = \left( \frac{\alpha_s C_F}{\pi}
\right)^4 \frac{\pi^3}{(N_c^2-1)^2} \left|
\int d^2k d^2k' f^Q(k,k',y) \right|^2.
\end{equation}
Here, and throughout the paper, momentum vectors are transverse momentum
two-vectors of positive norm. We follow Mueller and Tang \cite{mt} in
choosing the normalisation such that if $f^Q(k,k',y)$ is the amplitude for
scattering two gluons (transverse momenta $k$ and $k'$) then the cross section
of eq.(2) is that for quark-quark elastic scattering. The momentum
transfer $-t = Q^2$. The separation in rapidity between the final state
parton and meson is $y = \ln(\hat{s}/4\bar{q}^2)$ with $\hat{s} = x's$
($s$ is the $\gamma^* p$ CM energy). We define
$\bar{q}^2=q^2_{\|}+Q^2/4$ and $q^2_{\|}=(Q^2_{\gamma}+m^2_V)/4$
($-Q^2_{\gamma}$ is the photon virtuality).
We use $-t$ and $Q^2$ interchangeably throughout the paper but it should not
cause confusion.

For vector meson production, the amplitude $f^Q$ is the product of the
two-gluon scattering amplitude (see eq.(5)) and the form factor associated
with the $\gamma^*V$ vertex:
\begin{equation}
\psi_0^V = {\cal C} \left[ \frac 1{2\bar{q}^2}\,-\,
\frac 1{2q_{\|}^2+2(k-Q/2)^2}\right]
\end{equation}
where
$$ {\cal C}^2 = \frac{3\Gamma^V_{e^+e^-}m^3_V}{\alpha_{em}}.$$
This is the result of ref.\cite{ry} (which is valid for heavy vector
mesons, e.g. $J/\Psi$ and $\Upsilon$). It is written in terms of the electronic
width $(\Gamma^V_{e^+e^-})$ of the meson, so as to cancel a large part of
the $O(\alpha_s)$-corrections.
As usual, $\alpha_{em}=1/137, \; C_F=4/3$ and $C_A=N_c=3$. Although eq.(1) is
written for heavy photon-proton interactions, it is also valid in the
photoproduction limit, i.e. $Q^2_{\gamma}=0$. Also, eq.(1) is the
component of the cross section corresponding to the scattering of transversely
polarised photons only. The longitudinal contribution is obtained
using $\sigma_L = (Q_{\gamma}^2/M_V^2) \sigma_T$ and so it dominates at
large enough $Q_{\gamma}^2$. Recall that to get the electroproduction
($ep$) cross section one has to multiply eq.(1) by the photon flux, i.e.
$$ \frac{\alpha_{em}}{\pi} \frac{dx}{x} \frac{dQ^2_\gamma}{Q^2_\gamma}
(1-Y+Y^2/2)$$ where $x$ is Bjorken-$x$ and $Y$ is the photon energy fraction
$(=Q^2_\gamma/(4xE_eE_p))$.

For the case of quark-quark elastic scattering, the form factor of eq.(3) is
replaced by unity and the asymptotic BFKL behaviour of the cross section has
been calculated to be \cite{mt}
\begin{equation}
\frac{d\sigma(qq \to qq)}{dt}\;=\;(\alpha_sC_F)^4\frac{\pi^3}{4t^2}
\frac{\exp{(2\omega_0y)}}{[\frac 7{2}\alpha_sC_A\zeta (3)y]^3}.
\end{equation}
Where $\zeta (3)\simeq 1.202$ is the Riemann zeta function and
$\omega_0 = \frac{C_A \alpha_s}{\pi}4\ln2$. In the next section, we calculate
the cross section for diffractive vector meson production in the same
(asymptotic) limit. We obtain analytic expressions
in the limits $|t|\gg q^2_{\|}$ and $|t|\ll q^2_{\|}$. We compare the results
with those obtained in the Born approximation (i.e. the two-gluon exchange
model of Low and Nussinov \cite{ln}). These asymptotic expressions
are only valid for $\alpha_s y \gg 1$, i.e. the rapidity interval should be
larger than the inverse QCD coupling. For
smaller $y$, we expect significant deviations from the asymptotic formulae,
and in section 3 we look for such deviations. Specifically, we iterate the BFKL
kernel numerically, so as to obtain corrections to the Born term and to
investigate the convergence of the BFKL series (recall it is an expansion in
$\sim \alpha_s y$). We perform our calculations up to and including terms
$\sim (\alpha_s y)^3$. We also show how to improve the convergence of the
series by performing an all orders summation of the logarithms which are large
in the asymmetric configurations where one (or more) of the $t$-channel gluons
is nearly on shell. This resummation allows us
to make predictions for the diffractive cross section for larger values of
$y$ than if we had used the simplest expansion. Of course, for sufficiently
large $y$, even the resummed series is poorly convergent and a
full solution to the BFKL equation is required.

It should be acknowledged that, contrary to the case of elastic hadron-hadron
scattering \cite{bs}, the lowest-order photon-hadron amplitude is not
sensitive to Sudakov corrections. It was shown in ref.\cite{fsz} that
the Sudakov logarithms, which arise when internal quarks or gluons go on-shell,
do not occur due to the pointlike nature of the photon. In addition, due to the
largeness of $\bar{q}^2$, we need not worry about vector meson dominance
contributions to the vector meson production vertex. This is the case even in
the photoproduction limit of $Q_{\gamma}^2 = 0$.

In section 4 we return to the diagrams of figs.1(c,d).  Such diagrams
should be important at large $M^2 (\simeq Q^2(1/x'-1))$.
A summary of our results and predictions for
cross sections in the HERA range are presented in the conclusion.

\section{High energy asymptotics}
The solution for the QCD pomeron amplitude at non-zero momentum transfer was
given by Lipatov \cite{lip} in terms of the eigenfunctions in the transverse
coordinate ($\rho$) representation, i.e.
\begin{equation}
f^Q(k,k',y) = \frac 1{(2\pi )^6}
\int d\nu\frac{\nu^2}{(\nu^2+1/4)^2}\exp{[\omega
(\nu)y]}  I^{A*}_{\nu}(k',Q)I^B_{\nu}(k,Q),
\end{equation}
where
\begin{equation}
I^A_{\nu}(k,Q)\;=\;V^A(k,Q) \int d^2\rho_1d^2\rho_2\left
(\frac{(\rho_1-\rho_2)^2}{\rho^2_1\rho^2_2}\right )^{(1+2i\nu )/2}
{\mathrm e}^{[ik\cdot\rho_1+i(Q-k)\cdot\rho_2]}.
\end{equation}
The co-ordinate vectors of the $t$-channel gluons are $\rho_1$ and $\rho_2$
(they are the vectors conjugate to the transverse momentum vectors $k$ and
$k-Q$) and $V(k,Q)$ is the `impact factor' for the vertex, $A$.
For the coupling to a quark line, one can take an impact factor of unity with
the modification of eq.(6) which is appropriate for scattering off coloured
particles (this is the prescription of ref.\cite{mt} and we shall discuss it
in more detail in section 4). Whilst for the $\gamma^* V$
vertex we take the form factor of eq.(3). The function $\omega(\nu)$ is
$$ \omega(\nu) = \frac{2 \alpha_s C_A}{\pi} {\cal R}{\mathrm e}
[\psi(1)-\psi(1/2+i\nu)]
$$ with $\psi(z) = (d/dz)\ln \Gamma(z)$ (the derivative of the logarithm of the
gamma function).

In the high energy limit $y\to \infty$ and the dominant contribution comes
from the saddle point at $\nu = 0$.
Thus we are interested in the function $I^A_0$. In the $qq$
case it was calculated in ref.\cite{mt}:
\begin{equation}
I^q_0(Q) = \int \frac{d^2k}{(2\pi )^2}I^q_0(k,Q)\;=\;-\frac{4\pi}Q.
\end{equation}
In the $\gamma^* V$ case, things are a little more complicated. Let us define
\begin{eqnarray}
V(\rho ) &=& \int\frac{d^2k}{(2\pi )^2}{\mathrm e}^{i(k-Q/2)\cdot\rho}\left [
\frac 1{2\bar{q}^2}\,-\,\frac 1{2q_{\|}^2+2(k-Q/2)^2}\right ] \nonumber \\
&=& \frac 1{2\bar{q}^2} \delta^{(2)}(\rho )\,-\,\frac 1{4\pi}K_0(q_{\|}\rho),
\end{eqnarray}
$R=(\rho_1+\rho_2)/2, \; \rho =\rho_1-\rho_2$ and $K_0$ is the McDonald
function. Now instead of eq.(7), we have
\begin{eqnarray}
I^V_0(Q) &=& {\cal C} \int d^2Rd^2\rho\frac{|\rho|}{|R+\rho/2| |R-\rho/2|}
{\mathrm e}^{iQ\cdot R}V(\rho) \nonumber \\
&=&-{\cal C} \int J_0(QR) \; R\,dR \; \frac{|\rho|}{|R+\rho/2| |R-\rho/2|}
\frac{K_0(q_{\|}\rho)}2 d^2\rho
\end{eqnarray}
and $J_0(x)$ is the Bessel function of the first kind.
The cross section is then given by
\begin{eqnarray}
\frac{d\sigma^T(\gamma^*p\to V+X)}{dt dx'} &=& \left
( \frac{81}{16}G(x',t)+\sum_f(q(x',t)\,+\,\bar{q}(x',t))\right )
\frac{d\sigma(qq\to qq)}{dt} \nonumber \\ & & \times  \left[
I^V_0(Q) \left( \frac{4\pi}Q \right)^{-1} \right]^2,
\end{eqnarray}
where the quark-quark scattering cross section is given in eq.(4).
Eq.(9) can be evaluated analytically\footnote{We are indebted to Hans Lotter
for correcting a mistake in our initial evaluation of eq.(11)}
in the limits $Q^2\gg q^2_{\|}$ and $Q^2\ll q^2_{\|}$.

For $Q^2\gg q^2_{\|}\,$ (see ref.\cite{lot})
\begin{equation}
I_0^V(Q)\vert_{Q^2 \gg q^2_{\|}} \simeq - \frac{{\cal C}}{Q^3} \ln \left(
\frac{4 Q^2}{q_{\|}^2} \right) \frac{32 \pi^3}{\Gamma^4(1/4)}.
\end{equation}


In the other limit of small $Q^2\ll q^2_{\|}\,$, the essential contribution
comes from $R\gg \rho /2$. The logarithmic integral over $R$ goes from $\rho/2$
to $1/Q$ and is equal to $\int_{\rho /2}^{1/Q}\frac{dR}R \simeq
\ln \frac{2q_{\|}}Q$, while the integral
$\int^{\infty}_0 d^2\rho \rho \; K_0(\rho q_{\|})=\pi^2/q_{\|}^3$. So
\begin{equation}
I_0^V(Q)\vert_{Q^2\ll q^2_{\|}}\simeq -\frac{{\cal C}}{4q_{\|}^3}
\ln\left(\frac{4q_{\|}^2}{Q^2}\right) \pi^2.
\end{equation}

In fig.(2), the results of an exact numerical computation of $I^V_0$ (eq.(9))
are compared with the asymptotic formulae of eqs.(11) and (12), as a function
of the ratio $\tau \equiv Q^2/4q^2_{\|}$. This is a sensible variable to
plot all our results against, since it exploits the corresponding
scale invariance of the BFKL kernel.

To finish the section let us compare the asymptotic BFKL behaviour for
the QCD pomeron amplitude with that obtained in the Born approximation. For
the two-gluon pomeron the cross section for $\gamma^* q \to Vq$ takes the form
\begin{equation}
\frac{d\sigma^T(\gamma^* q \to Vq)}{dt} = {\cal C}^2
\frac{4\pi \alpha_s^4}{81} {\cal J}^2,
\end{equation}
where
\begin{eqnarray}
{\cal J} &=& \frac 1{\pi}\int \frac{d^2k}{k^2(Q-k)^2}
\left [ \frac 1{2\bar{q}^2}\,-\,\frac 1{2q_{\|}^2+2(k-Q/2)^2}\right ]
\nonumber \\ &=&
\frac 8{Q^4-16q^4_{\|}}\ln\frac{(Q^2+4q^2_{\|})^2}{16Q^2q^2_{\|}}.
\end{eqnarray}
For large $Q^2\gg q^2_{\|}\,$ the leading
logarithmic contribution to ${\cal J}$ comes
from $q^2_{\|}\ll(k-Q/2)^2\ll Q^2/4$ and ${\cal J}=\frac 8{Q^4}
\ln\frac {Q^2}{16q^2_{\|}}$.
The same logarithmic behaviour arises in the asymptotic BFKL calculation
(eq.(11)), which contains an additional enhancement due to the factor $\sim
\exp{(2\omega_0y)}/y^3$. Note that at not too large energies, the numerical
coefficient of the BFKL asymptotic result can be small and for
$\omega_0 y \approx 1.5$ it is only about $3\%$ of the Born term.
%
%

For small $Q^2 \ll q^2_{\|}\,$
%
%
the BFKL factor of $$\frac{\pi^4}{4Q^2}
\ln^2\frac{4q_{\|}^2}{Q^2} \exp{(2\omega_0y)}/
[\frac 7{2}\alpha_sC_A\zeta (3)y]^3$$ is to be compared with the
$$\frac 1{q^2_{\|}}\ln^2\frac{q^2_{\|}}{Q^2}$$ obtained
in the Born approximation. There is an additional enhancement of
$\sim q_{\|}^2/Q^2$ in the case of the BFKL pomeron. The origin of the
difference is clear. The distance between two gluons in the Low-Nussinov
pomeron is fixed since they must couple to the $q\bar{q}$ pair, i.e. their
size does not exceed that of the upper vertex ($\rho\sim 1/q_{\|}$). Thus the
two-gluon system is not sensitive to small $Q^2$. However, in the BFKL
pomeron, after a few iterations (or rungs of the ladder) the gluons can be
separated by large distances $\rho\sim 1/Q$ and so the amplitude $\sim 1/Q$.

\section{Iterations of BFKL kernel}
So far, we have considered only the asymptotic solution to the BFKL equation.
In this section we consider order-by-order iteration of the BFKL kernel for
non-zero $t$ and examine the nature of the leading log $s$ expansion. For
sufficiently small $z$ ($= y C_A\alpha_s/2\pi)$ (and sufficiently large $|t|$)
we can expect the Born approximation to be appropriate. As $z$ increases, so
the need to include more and more terms in the BFKL series increases until the
point is reached where an all orders summation is vital. By studying the BFKL
series in this way we can hope to improve on the Born cross section estimate
in the region of intermediate $z$. In addition, such an expansion will allow us
to identify the onset of the region where the BFKL (all orders) resummation is
vital. To start, let us recall the BFKL equation
for non-zero $t$:
\begin{eqnarray}
\Phi_i(k_i,z_i) = &  & \frac{1}{\pi} \int_0^{z_i} dz_{i-1}
\int \frac{d^2k_{i-1}}{(k_i-k_{i-1})^2} \times  \\ & & \left\{
\Phi_{i-1}(k_{i-1},z_{i-1}) \left[ \frac{\hat{k}_{i}^2}{\hat{k}_{i-1}^2} +
\frac{k_{i}^2}{k_{i-1}^2} -
Q^2 \frac{(k_i-k_{i-1})^2}{k_{i-1}^2 \hat{k}_{i-1}^2} \right]
\right. \nonumber \\ & & \left. -\Phi_{i-1}(k_i,z_{i-1}) \left[
\frac{k_i^2}{k_{i-1}^2 + (k_i-k_{i-1})^2} +
\frac{\hat{k}_i^2}{\hat{k}_{i-1}^2 + (k_i-k_{i-1})^2} \right] \right\}
\nonumber
\end{eqnarray}
For iterations from the quark line we use the boundary condition:
\begin{equation}
\Phi_0(k,z) = 1.
\end{equation}
As usual, all the $k_i$ are two-vectors in the transverse plane and $\hat{k} =
k-Q$. The first iteration of this input leads to the simple expression:
\begin{equation}
\Phi_1(k,z) = z \ln \left(
\frac{k^2\hat{k}^2}{Q^4} \right).
\end{equation}
Subsequent iterations are difficult to perform analytically and we evaluate
them numerically.

Successive iterations lead to large oscillatory behaviour in the regions
$k^2 \ll Q^2$ and $\hat{k}^2 \ll Q^2$ (as a result of logarithms like the one
in eq.(17)).  This behaviour leads to a very poorly convergent
series. To overcome this difficulty we will sum up the (double)
logarithmic contributions $\sim[z \ln(k^2/Q^2)]^n$ analytically, and then
factorise these badly oscillating terms to leave behind a better convergent
series.

There are no infrared divergences in eq.(15) (neither at $k'\to k$,
nor at $k'\to 0$ or $k'\to Q$) and the only logarithm comes from the last
(reggeization) term in the region $k^2\ll k'^2 \ll Q^2$.
For small $k^2\ll Q^2$ (or $\hat{k}^2 \ll Q^2$) eq.(15) leads to
$$\frac{\partial \Phi(k,z)}{\partial z}\,=\,\frac 1{\pi} \int\frac{d^2k'}
{(k'-k)^2}\left\{ \left [-\frac{(k'-k)^2}{k'^2}+1+\frac{k^2}{k'^2}\right ]
\Phi(k',z)\right.$$
\begin{equation}
\left. -\,\Phi(k,z)\left [\frac{k^2}{k'^2+(k'-k)^2}+1\right ]\right \},
\end{equation}
and we have introduced the function $\Phi(k,z)$, where
$$\Phi(k,z) = \sum_{n=0}^{n=\infty} \Phi_n(k,z).$$
We have used the result that $\frac{\partial \Phi_0(k,z)}{\partial z} =0$.
After the angular integration
$$\frac{\partial \Phi(k,z)}{\partial z}\,= \int dk'^2\left \{\left
[-\frac 1{k'^2}+ \frac 1{|k'^2-k^2|}\left (1+\frac{k^2}{k'^2}\right )\right ]
\Phi(k',z)\right.$$
\begin{equation}
\left. -\,\Phi(k,z)\left[\frac{1+k^2/k'^2}{|k'^2-k^2|}-
\frac{k^2/k'^2}{\sqrt{4k'^4+k^4}}\right ]\right \}.
\end{equation}
Which can be written more concisely:
\begin{equation}
\frac{\partial \Phi(k,z)}{\partial z} =
\int \frac{d\xi '}{\xi '}\left \{\left[\frac{\xi +\xi '} {|\xi -\xi '|}-1
\right] (\Phi(k',z)-\Phi(k,z)) + \Phi(k,z)\left[
\frac 1{\sqrt{\frac{4\xi'^2}{\xi^2}+1}}-1\right]\right\},
\end{equation}
where $\xi=k^2/Q^2$ and $\xi'=k'^2/Q^2$.
It is easy to see that the only logarithmically large contribution comes from
the last term, which gives\footnote{At first sight the
accuracy of eq.(18) is insufficient to be sure we are not missing some constant
in comparison with the $\ln\xi$. However this log comes only from the
reggeization part of the BFKL kernel, for which the exact answer is known. The
contribution is equal to $\Phi(k,z)[\ln(Q^2/\mu^2)+\ln(k^2/\mu^2)]$. The
infrared cutoff $\mu^2$ is cancelled after we subtract the part $$2\Phi(k,z)
\left[\int^{\xi}_0\frac{d\xi '}{\xi -\xi '}\,+\,\int^{\infty}_{\xi}
\frac{d\xi ' \xi}{\xi '(\xi '-\xi)}\right]\,=\,2\Phi(k,z)\ln(k^2/\mu^2),$$
which has already been included in  the integral of eq.(21).
Finally one gets $\Phi(k,z)\ln\xi $ and the accuracy of eq.(20) is of the
order of $O(k^2/Q^2)$.} $\Phi(k,z)\ln\xi $ and so
$$\frac{\partial \Phi(k,z)}{\partial z}\,=\,2\left[
\int^{\xi}_0\frac{d\xi' (\Phi(k',z)-\Phi(k,z))}{\xi -\xi '}\right.$$
\begin{equation}
\left. +\,\int^{\infty}_{\xi}\frac{d\xi'
(\Phi(k',z)-\Phi(k,z))\xi}{\xi '(\xi '-\xi)}\right]\,+\,\Phi(k,z)\ln\xi.
\end{equation}
In the double log approximation the solution is
$$\Phi(k,z)=\exp{(z\ln\xi)}$$
(the integral over $\xi'$ does not give rise to any logs). Now we can put
$\Phi(k,z)=\phi (z,\xi )\exp{( z\ln\xi )}$ and
compute the solution for $\phi$
$$\frac{\partial \phi (z,\xi )}{\partial z}\,=\,2\left[ \int^{\xi}_0
\frac{d\xi' (\phi (z,\xi ')(\xi '/\xi )^z-\phi (z,\xi ))}{\xi -\xi '}\right.$$
\begin{equation}
\left. +\,\int^{\infty}_{\xi}\frac{d\xi' (\phi (z,\xi')(\xi'/\xi )^z-
\phi(z,\xi ))\xi}{\xi'(\xi'-\xi)}\right],
\end{equation}
with the initial condition $\phi(0,\xi )=1$ (corresponding to iterations from
the quark line). However one can only make sense of such a framework for
$z \lapproxeq 1$. At larger $z$ the last integral in eq.(22) diverges in the
region of $\xi' \to\infty$. The implication is that, for $z \gapproxeq 1$,
eqs.(18-22) are not self-consistent,
i.e. the main contribution to $\Phi(k,z)$ (with small $\xi$)
comes from the region of large $\xi'$. The results of the
numerical calculation are presented in fig.3. For $z \lapproxeq 0.4$ the
variations of the function $\phi$ (from unity) are rather small.

Let us explain how the diffractive cross section is obtained from the
$\Phi_m(k,z)$ functions. First define (and we now make explicit the dependence
upon the scaling variable $\tau =Q^2/4q^2_{||}$)
\begin{equation}
\psi(z,\tau) = \sum_{n=0}^{n=\infty}
\int \frac{d^2k}{\pi} \frac{\psi^{q}_{m}(k)
\psi^{V}_{n-m}(k)}{k^2 \hat{k}^2}
\frac{z^n}{n!}
\end{equation}
where $\psi^q_m(k) z^m/m! = \Phi_m(k,z)$ is a dimensionless
function corresponding to $m$ iterations of the BFKL kernel starting from the
quark line boundary condition, $\psi^q_0(k) = 1$. Iterations from the
vector meson boundary condition of eq.(3) lead to the functions $\psi_n^V(k)$.
The Born term is just the lowest order convolution of eq.(14). The Born term
and first iteration from the quark line are known analytically, and the second
iteration has been computed numerically. Similarly, the zeroth iteration from
the vector meson boundary condition is known analytically, whilst the first
iteration can be computed numerically etc.. Consequently, we can convolute
these
results to obtain corrections to the Born contribution. Where possible we have
checked our results by calculating the convolutions of eq.(23) in different
ways, i.e. we have checked that $\psi^q_1 \otimes \psi^V_1 = \psi^q_2 \otimes
\psi^V_0$. For the cross
section we have
\begin{equation}
\frac{d\sigma^T(\gamma^* q \to Vq)}{dt} = \frac{4\pi\alpha_s^4}{81}
|\psi(z,\tau)|^2.
\end{equation}
After summing the double logarithms
$$ \sum_{n=0}^{\infty} \psi^q_n(k) \frac{z^n}{n!} =
\exp(z\ln \xi \hat{\xi}) \sum_{n=0}^{\infty}
\phi^q_n(k) \frac{z^n}{n!} $$
and the equality means that
$$
\phi^q_n(k) = \psi^q_n(k) - \psi^q_{n-1} \ln \xi\hat{\xi} \frac{n!}{(n-1)!} +
....... + \psi^q_{n-m}(k) \ln^m \xi \hat{\xi}\frac{(-1)^m n!}{(n-m)!m!} +
.......
$$
We choose to resum terms $\sim z \ln \xi \hat{\xi}$ (this expression
simultaneously reduces to the double log solution when $\xi \ll 1$ or
$\hat{\xi} \ll 1$). By convoluting the $\phi^q_n$ functions (rather than the
$\psi^q_n$) with $\psi^V_0$, in eq.(23), we can compute the
double log improved cross section. It should
be appreciated that it is only meaningful to perform the exponentiation in the
small $\xi \hat{\xi}$ region. Since the integrals (in the convolutions)
are performed over all $\xi$ (up to infinity), we must introduce a
`factorisation' scale, $\xi_0$, to delineate the small $\xi \hat{\xi}$
region, i.e. $\xi \hat{\xi} \leq \xi_0$ must be satisfied before
resumming. Choosing $\xi_0=0$ therefore corresponds to the un-resummed series.
We choose $\xi_0 = 1$  but the choice is essentially arbitrary (in the same
way that the QCD factorisation scale is arbitrary). Essentially, the larger
one chooses $\xi_0$, the better the effect of resummation -- up to the point
where the logarithm of $\xi_0$ can no longer be considered `large'.

In figures 4 to 6 we show the dependence of the function
$\psi(z,\tau)$ upon $\tau$ for different values of $z$ (in
fact we plot $[\psi(z,\tau) t^2 /(2 {\cal C})]^2$).
These are essentially plots of the cross section, which can
be obtained from the plots via eq.(24) and eq.(1). Notice that we have removed
the explicit $1/t^4$ behaviour of the cross section to leave behind a quantity
which only depends upon the `scaling' variable, $\tau$, and
the `energy' variable, $z$. In all cases, the dotted lines are the
Born\footnote{Or the Born plus the double logs in the case of fig.(5,6).}
results, the dashed lines include the 1st ($\sim z$) BFKL corrections
(to the Born results), the solid lines include the 2nd ($\sim z^2$) BFKL
corrections and the diamonds include the 3rd ($\sim z^3$) BFKL corrections.
The dash-dotted lines are obtained using the
asymptotic results of the previous section (i.e. the numerical evaluation of
eq.(9)). In fig.(4), we show results for the simplest expansion of the BFKL
series, i.e. without the double log resummation. Fig.4(a) shows quite clearly
the good convergence that is expected for small enough $z$ ($z = 0.05$).
The Born term provides a reasonable description  and lies within
$30\%$ of the higher order result over most of the $\tau$ range. The
major deviations arise around the dip near $\tau = 1$ (the
cross section is zero here in the Born approximation due to the
cancellation between the diagrams corresponding to the two gluons
coupling to the same and different quark lines). Going to $z=0.35$,
as we do in fig.4(b), we find very poor convergence for $\tau \gapproxeq 1$
(where the 2nd and 3rd iterations often deviate by
an order of magnitude). The problem is a result of the
oscillatory nature of the solutions to the BFKL equation. The first
corrections to the Born term are big and they completely fill in the
dip. The next corrections tend to restore the dip, by including large
negative contributions in the region of $\tau \gapproxeq 1$
which partially cancel the positive contribution of the first
corrections. This leads to the unpredictable behaviour of fig.4(b). We do not
expect, therefore, to make reliable predictions for the diffractive cross
section using such a simple minded expansion of the BFKL series (for all but
the smallest values of $z$). Before moving on to the results obtained with the
resummed series we should point out that, as noted above, the poorest
convergence of the un-resummed series occurs for the larger values of $\tau$,
(in particular in the region around the dip) and that this is a general
feature of our results. Indeed we shall see that it is precisely this region
that is improved by resumming. This is to be understood as the region where the
asymmetric configuration (i.e. where one of the gluons which couples to the
$\gamma V$ vertex carries all the momentum transfer) is important.

The double log improved results are shown next. Fig.5 shows the analogous
plots to those in fig.4. The accuracy of the (resummed) Born term is improved
at $z=0.05$, whilst at $z=0.35$ the higher corrections are necessary. However,
for $z=0.35$, we can now claim to make reasonable predictions over the whole
$\tau$ range after just two iterations of the BFKL kernel, i.e. the
difference between the 2nd and 3rd iterations is typically $\lapproxeq 20 \%$.
Fig.6 shows the resummed results for $z=0.2, 0.5$ and $0.8$ (which are typical
values for HERA). For $z=0.2$ we obtain convergence at the level of $\sim 1\%$
at order $z^3$. Going to larger $z$, as one expects, the quality of convergence
deteriorates and, at $z=0.8$, we clearly need a more complete summation of the
BFKL series. However, it is essential to realise (since the cross section
is expected to vary rapidly with the momentum transfer) that a prediction of
the cross section to within a factor $\sim 2$ is very useful. For example,
at $z=0.8$, the Born cross section lies typically more than an order
of magnitude ($\sim 50$) below
the order $z^3$ results (diamonds), which is in turn a factor $\sim 2-3$ below
the asymptotic results.

Comparison of the numerical results obtained in this section with the
asymptotic results (dash-dot lines) shows quite clearly that the asymptotic
formulae do not represent the diffractive cross section over most,
if not all, of the kinematic range examined. Only at the largest value of $z$,
does it look likely that, after fully summing the BFKL series, the result
will agree with the asymptotic prediction. In the region where the double
logarithmic resummation is most important, i.e. $\tau \gapproxeq 1$, the
asymptotic formulae fail completely to approximate even the gross features of
the expected shape, e.g. the dip. In addition, the asymptotic prediction always
appears to overestimate the size of the cross section. Ultimately, for large
enough $z$, the asymptotic result must be more appropriate than the `fixed
order' results. We can see, in fig.6, that at $z=0.8$ the BFKL
series is slowly convergent and one can no longer trust the order $z^3$
predictions, i.e. this marks the onset of the dynamics manifest in the full
BFKL solution which will ultimately be described by the asymptotic solution.
It should not come as a surprise that the asymptotic prediction is
not trustworthy for $z \lapproxeq 1$. Its reliability relies upon the validity
of the approximation that the exponential in eq.(5) can be expanded about $\nu
= 0$. This is a very poor approximation for $z \lapproxeq 1$ since the
integrand of eq.(5) actually vanishes at $\nu = 0$ and rises rapidly to its
maximum value only for large $z$.

It is also worth commenting upon the results of Lipatov \cite{lip} and of
Hancock and Ross \cite{hr}. They have found that, in the case of a running
$\alpha_s({\mathrm {Max}}(k^2,Q^2))$,
the leading eigenvalue of the BFKL kernel lies somewhat below its
asymptotic value for all realistic energies (and non-zero $t$). Therefore the
cross section is expected be smaller than that predicted by the asymptotic
solution based on eq.(4). We do not investigate the effect of a running
coupling here.

\section{Sub-leading Corrections}
We have devoted a whole section to the calculation of the process
$\gamma^* q \to V q$. We now turn to a discussion of the `lower part' of the
diagram and the validity of the simple assumption that the pomeron couples to a
single parton inside the proton.

If $Q^2$ is large enough, so that the partons (which we label by their
longitudinal momentum fractions $x'$ and $x''$) within the proton are
separated by distances $\rho'\gg 1/Q$ then we are certainly entitled
to restrict ourselves to the simple coupling to a single parton line.
This may seem somewhat suprising, since at first sight eq.(6) appears to
suggest that the only contribution arises from the coupling to different
parton lines (the amplitude vanishes at $\rho_1 = \rho_2$): This is not the
case. To exploit the conformal invariance, Lipatov redefined the eigenfuctions
(by adding terms proportional to $\delta^{(2)}(k)$ and $\delta^{(2)}(k-q)$
which
vanish on coupling to colourless particles) and so the contribution from the
Feynman graphs where the pomeron couples to a single parton is hidden. Indeed,
let us start by assuming that $\rho' \gg 1/Q$, such that (for large $Q^2$)
\begin{equation}
\int d^2 R\; {\mathrm{e}}^{iQ\cdot R} \frac{\rho}{| R + \rho/2| | R - \rho/2|}
\approx \frac{2\pi}{Q} ( {\mathrm{e}}^{iQ\cdot \rho/2} +
{\mathrm{e}}^{-iQ\cdot\rho/2} ),
\end{equation}
i.e. the main contributions arise from the regions where $R$ is within
$\Delta R \sim 1/Q$ of the singular points at $\pm \rho/2$. This is the
result which underpins the Mueller-Tang prescription for
calculating the contribution from the coupling to a single parton line. The
first term of the r.h.s. of eq.(25)
corresponds to interaction with a quark at $\rho' =
\rho/2$ whilst the second term is due to interaction with an anti-quark at
$-\rho/2$. Corrections to eq.(25) are $\sim (\Delta R/\rho)^2 \sim
1/(\rho^2 Q^2)$ and represent contributions from Feynman graphs where the
pomeron couples to more than one parton.

Note that this approach is not appropriate for the $\gamma V$ vertex. In
evaluating the $R$ integral of eq.(9) we are not entitled to assume that the
essential contributions arise from the regions around the singular points since
the subsequent $\rho$ integral would be dominated by contributions from small
$\rho \sim 1/Q$, i.e. after evaluating the $R$ integral using eq.(25), we would
be left with $$ \int K_0(q_{||}\rho) {\mathrm{e}}^{iQ\cdot \rho/2} d^2\rho $$
where the dominant contribution arises from small $\rho \sim 1/Q$ and thus we
are not justified in assuming that $\rho \gg \Delta R$.
The corrections are therefore essential and indeed give the main contribution
to $I^V_0(Q)$ (i.e. the logarithm in eq.(11)) at large $Q^2$.

At the lower vertex, the situation is quite different since the target
dissociates. The impact factor now takes the form,
\begin{equation}
|I^p_0(Q)|^2 = 2 \int d^2\rho\; d^2 R \; d^2 R'\; |\Psi(\rho)|^2
\frac{\rho\; {\mathrm{e}}^{iQ\cdot  R }}{|R-\rho/2||R+\rho/2|}
\frac{\rho\; {\mathrm{e}}^{-iQ\cdot R'}}{|R'-\rho/2||R'+\rho/2|}
\end{equation}
where $|\Psi(\rho)|^2$ is analogous to $V(\rho)$ in eq.(8).
The integrals over
$R$ and $R'$ can now be performed safely in the $\rho \gg \Delta R$ limit since
the exponentials (${\mathrm{e}}^{iQ\cdot \rho/2}$ and
${\mathrm{e}}^{-iQ\cdot \rho/2}$) cancel each other. Thus we get
\begin{equation}
|I^p_0(Q)|^2 = \frac{16 \pi^2}{Q^2} \int d^2 \rho |\Psi(\rho)|^2.
\end{equation}
With the lower limit of the $\rho$ integral equal to $1/Q$ and the upper
limit fixed by the proton radius the integral over $\Psi$ gives the
conventional structure function, i.e. $|\Psi|^2 \sim 1/\rho^2$ generates the
leading logarithmic behaviour of the parton density function with the typical
values of $\rho \gg 1/Q$. Since the integrand of the $\rho$ integral is
independent of $Q^2$ (it is simply the wavefunction) it follows that the main
term in eq.(26)
is indeed the contribution from the graphs where the pomeron couples to a
single parton line. Thus we have obtained the Mueller-Tang result
that was used in the first part of the present paper and have justified our
assumption that the dominant contribution to the proton dissociation (at
large enough $Q^2$) arises from the coupling to a single parton line.
For a more detailed discussion of the nature of the pomeron-quark coupling we
refer to ref.\cite{bart}.

Note that we neglected the cross terms $\sim {\mathrm{e}}^{iQ\cdot \rho}$ which
arise  from the points $R\to \pm\rho /2;\;\;\; R'\to \pm\rho /2$
after the $R$ and $R'$ integrals of eq.(26). They represent contributions
from the interference between graphs where the pomeron couples to single but
different parton lines. Such contributions are suppressed at least by
logarithms of $Q^2$ since the momentum transfer must flow into the proton
wavefunction.

Similarly, diagrams like the one of fig.1(c) are beyond the leading $\ln Q^2$
approximation since they lose at least one power of $\ln Q^2$ due to the
hooking of the large transverse momentum, $k \sim Q/2$ into lower rungs of the
`structure function'.\footnote{It was shown in section 2 that for
large $y$ the internal momenta inside the
BFKL ladder are typically $k\sim Q/2$. The contribution
from the region of small $k$ (or $(Q-k)$) dies out with energy. Even in the
Born (two-gluon) approximation the main logarithmic contribution comes from
$(k-Q/2) \ll Q/2\,$, i.e. $k$ close to $Q/2$.}
Therefore the contributions from such graphs are suppressed
in comparison with those of figs.1(a,b) by the factor $-\frac{1}2\gamma_2$. The
anomalous dimension of the twist-2 operator (structure function) is
$\gamma_2\propto \alpha_s$ and its presence reflects the fact that one loses
one power of $\ln Q^2$. The additional smallness $(-1/2)$ comes from the colour
coefficient ($C_A/2$ in comparison with $C_A$). Note that the coefficient is
negative and so the graph in fig.1(c) describes a colour screening
effect. If the BFKL gluon continues down and touches the next $s$-channel
gluon (i.e. $x'''$) then one gets a factor of $1/4$ and two powers of
$\gamma_2$, and so on. When computing the cross section, one must remember to
double the contributions from configurations where the two $t$-channel gluons
couple to different $s$-channel gluons. Finally, if the two gluons couple to
the same $s$-channel gluon $(x'')$ then one loses one logarithm and picks up
a colour coefficient which is the square of the result obtained when the
gluons couple to different gluons. Thus the leading contribution to the
order $\alpha_s$ corrections is suppressed relative to the contributions of
figs.1(a,b) by the factor
\begin{equation}
T_{2}\;=\;\left (-\frac{1}2-\frac{1}2+\frac{1}4\right)\gamma_2=-\frac{3}4
\gamma_2\simeq -0.3.
\end{equation}
We put $\alpha_s=0.16$ and use the double logarithmic approximation
result: $\gamma_2=C_A\alpha_s/(\pi\omega )\simeq 0.36$ with
$\omega =\omega_0=0.42$.
Such contributions correspond to the $\sim 1/(\rho^2 Q^2)$ corrections to the
impact factor, $I^p_0(Q)$, calculated above.

Let us now turn to graphs like the one shown in fig.1(d). Such diagrams differ
from those of fig.1(c) in the sense that one of the gluons now couples deep
inside the proton (e.g. to a different branch of the parton cascade). As such
we expect a power like suppression combined with an enhancement due to the
sampling of the two-parton component of the hadron wavefunction. Let us
estimate the possible size of such contributions. The contribution is
analogous to the graph of fig.1(b), in the sense that the pomeron is coupling
to different parton lines in the scattered object. Unlike the case of fig.1(b)
however, its contribution is not calculable in perturbation theory since it
is dependent upon the rather complicated structure of the proton wavefunction.
Since the proton is more diffuse than the compact quark-antiquark
pair (of the upper vertex) we anticipate that fig.1(d) will be suppressed at
all values of the momentum transfer considered here, i.e.
$Q \gg \Lambda_{QCD}$.

In comparison with eq.(1), the contribution of the graph of fig.1(d) to the
cross section contains an additional colour factor $1/(N_c^2-1)=1/8$.
There is additional suppression from the integration
over the transverse momenta of the $t$-channel gluons, $k_1$ and $k_2
(= Q-k_1)$, which connect the pomeron and the proton. In the case of
fig.1(a,b) there were no correlations between the momenta ($k_1$ , $k_2$) in
the amplitude $A$ and the corresponding momenta ($k'_1$ , $k'_2$) in the
complex conjugated amplitude $A^*$. One therefore has independent integrals
over $k_1$ and $k'_1$ which give two large  (of the order of $Q^2$)
factors. This is not the case for the diagram of fig.1(d). The transverse
momentum, $k_2$, is typically carried away by its associated final state gluon
which means that $k_2 \sim k_2'$ and we get only one large integral which is
$\sim Q^2$. The remaining integral is then over the intrinsic gluon momentum,
$\delta k \; (k'_2=k_2+\delta k)$ and we become sensitive to non-perturbative
effects in the proton, i.e. the momentum transferred over the $k'_2$  loop in
fig.1(d) is limited by the wave function of the target nucleon.
However, there is an enhancement due to the larger combinatorial factor
associated with coupling the two-gluons to different parton lines.

We can be more explicit by returning to the impact parameter representation.
As we have just shown, the square of the impact factor for coupling the
pomeron to a single parton line is $\propto (4 \pi/Q)^2$ (eq.(7,27)) and to
obtain the cross section, we then multiplied by the parton density factor
(which can be thought of as the multiplicity associated with the
pomeron-proton impact factor). An analogous, but slightly more
involved, calculation can be performed for the case of the diagrams like that
in fig.1(d), i.e. we can compute the square of the impact factor and multiply
by the colour factor (1/8) and the square of the parton densities to obtain the
cross section. The impact factor can be written thus:
\begin{eqnarray}
|\Delta I^{p}_0(Q)|^2 &=&2 \int d^2\rho\; d^2 R \; d^2 R'\; |\Psi_2(\rho)|^2
\times \nonumber \\ & &
\left[ \frac{\rho\; {\mathrm{e}}^{iQ\cdot  R }}{|R-\rho/2||R+\rho/2|}
\frac{\rho\; {\mathrm{e}}^{-iQ\cdot R'}}{|R'-\rho/2||R'+\rho/2|}-
{\mathrm{M.T.}}\right].
\end{eqnarray}
The Mueller-Tang subtraction term (denoted by `M.T.') is defined by the
replacement:
$$
\int d^2 R \frac{\rho\; {\mathrm{e}}^{iQ\cdot  R }}{|R-\rho/2||R+\rho/2|} \to
\int d^2 R \frac{\rho\; {\mathrm{e}}^{iQ\cdot  R }}{|R-\rho/2||R+\rho/2|}
- \frac{2\pi}{Q}({\mathrm{e}}^{iQ\cdot \rho/2} + {\mathrm{e}}^{iQ\cdot \rho/2})
$$
and similarly for the $R'$ integral. In this way we remove the contribution
corresponding to the coupling to a single parton line.
The answer, as it must be, is sensitive to the large distance
physics of the nucleon and this is contained in the two-particle wavefunction
$\Psi_2(\rho)$. Dependence upon longitudinal momenta is assumed to factorise
into the parton number densities.

We use a gaussian form for the $\rho$ distribution and can introduce the
same scale, $Q_0^2$, that is used by L.V. Gribov, Levin
and Ryskin (GLR) to parameterise the screening corrections to the deep
inelastic structure functions \cite{glr,lr}, i.e.
\begin{equation}
|\Psi_2(\rho)|^2 = \frac{Q_0^2}{4 \pi} {\mathrm e}^{- \rho^2 Q_0^2/4}
\end{equation}
with $Q_0^2 = 1.2$ GeV$^2$ from fits to $Sp\bar{p}S$ data (mainly on the
inclusive cross section for charged hadron production).

Eq.(29) can be simplified, i.e.
\begin{equation}
|\Delta I^{p}_0(Q)|^2 = 2 \int d^2 \rho\;  \rho^2 |\Psi_2(\rho)|^2 \; J_x(\rho)
J_y(\rho)
\end{equation}
where
\begin{eqnarray}
\frac{J_x(\rho)}{2} {\mathrm{e}}^{i Q\cdot \rho/2} &=& \int_0^{1/2} dx \left\{
\frac{K_0(Q\rho\sqrt{x(1-x)})}{\sqrt{x(1-x)}}({\mathrm{e}}^{ix Q\cdot \rho} +
{\mathrm{e}}^{i (1-x) Q\cdot \rho}) \right. \nonumber \\ &-& \left.
\frac{K_0(Q\rho\sqrt{x})}{\sqrt{x}} ( 1 + {\mathrm{e}}^{i Q\cdot \rho} )
\right\} \nonumber \\ &-& \int_{1/2}^{\infty} dx
\frac{K_0(Q\rho\sqrt{x})}{\sqrt{x}} ( 1 + {\mathrm{e}}^{i Q\cdot \rho} ).
\end{eqnarray}
and similarly for $J_y(\rho)$.
In the limit $Q/Q_0 \gg 1$ the integral is dominated by the contribution from
$x \sim 1/(Q\rho)^2 \ll 1$ and $\rho \sim 1/Q_0$. We can thus neglect the
integral over $x > 1/2$. After expanding the exponentials and performing the
remaining integrals we obtain:
\begin{equation}
|\Delta I^{p}_0(Q)|^2 =  \frac{16 \pi^2}{Q^2} \frac{Q_0^2}{4 Q^2} S
\end{equation}
with $S \sim \ln(Q/Q_0)$. We also computed $S$ numerically, and for
$4Q^2/Q^2_0=10$ we get $S=0.74$.

Combining eq.(33) with the colour factor and the square
of the parton densities we can now estimate the size of the
corrections which originate from fig.1(d), i.e. we have, rather than the
factor $G(x',t)dx'$, the term
\begin{equation}
\frac{S}{8} \left(\frac{Q^2_0}{4 Q^2}\right)\, G(x_1,Q^2/4)G
(x_2,Q^2/4) dx_1 dx_2.
\end{equation}

Now let us talk a little more about the value of $Q_0$. At first sight,
one might anticipate that $Q_0\sim 1/R_N$
(the inverse proton radius). However from the
semihard phenomenology of ref.\cite{lr} we know that the gluon-gluon
correlation length $R_0$ is much smaller than $R_N$. Also, calculations based
upon QCD sum rules determine the radius of the two-gluon form factor of the
proton to be $R_0 \simeq 0.3-0.35$ fm \cite{bgms}, which corresponds to
$Q^2_0 \simeq 1.2 \mbox{ GeV}^2$. We should point out that we have ignored
corrections due to interactions between the two branches of the parton
cascade (i.e. we used the square of the single parton density in eq.(34)).

We are ready to make a numerical estimate for the size of this correction.
The integrations over $x_1$ and $x_2$, in eq.(34), are limited  by the
mass of the hadron system $M^2\simeq \frac{|t|}4 \left(\frac 1{x_1} +
\frac 1{x_2}\right)$ or by other experimental conditions.
For example, one might choose to search for jets in the proton dissociation.
This would be a useful search to perform, since it would not only give an
additional way to measure the momentum transfer (and the momentum fractions
$x_1$ , $x_2$ carried by the jets) but it could be used as a means to
distinguish between the leading and sub-leading contributions (for the
sub-leading contribution, the vector meson transverse momentum is balanced by
a pair of jets each with transverse momentum $\sim Q/2$, while in the leading
case it is balanced by the one jet). Unfortunately,
it is not easy to make such a study at HERA since the need to ensure a large
rapidity gap forces the jets to lie close to the beam hole. Since we do not
require the observation of the proton dissociation, we shall estimate the
sub-leading corrections assuming no jets are seen in the proton direction, i.e.
$$\theta \simeq \frac{\sqrt{|t|}/2}{xE_p}< 3^o$$ in HERA lab frame then
$$ x_{1,2}> x_m\;=\;\frac{\sqrt{|t|}}{2E_p\theta} \approx 0.02 $$
for $|t|=3 \mbox{ GeV}^2$. Thus we find the ratio of the leading to
sub-leading contributions to be
\begin{eqnarray}
&\approx& \frac{
\frac{S}{8} \left( \frac{Q^2_0}{4 Q^2} \right) \int^1_ {x_m}
G(x_1, Q^2/4) dx_1  G(x_2, Q^2/4) dx_2}{\int^1_{x_m} G(x',t)dx'}
\nonumber \\
&\approx& \frac{S}{8} \left(\frac{Q_0^2}{4 Q^2}\right)
\int^1_{x_m}G(x,Q^2/4)dx\approx \frac{0.18S \mbox{ GeV}^2}{|t|}
\end{eqnarray}
and we have put $x_m=0.02$. Taking $Q^2=1.2$ GeV$^2$ and
$|t| \gapproxeq  3$ GeV$^2$ (i.e. $S\approx 0.74$) it follows that this ratio
is $\lapproxeq 5\%$.

As we mentioned above, perhaps the best way to study the sub-leading
contributions would be to look at the jet structure of the final state. At high
enough centre of mass energies one can have a large diffracted mass and still
maintain a rapidity gap. Events with a large diffracted mass will lead to
relatively large values for the sub-leading corrections (since $x_1$ and $x_2$
can be small).

Based on these estimates we think that, over the HERA range, the subleading
corrections discussed here do not change the results of the previous
calculations (which assume factorisation of the proton dissociation)
in a crucial manner. The effect may be less than $\approx 5\%$ for
dissociation into a small mass, but it would be interesting to study
experimentally how it reveals itself at higher masses.

\section{Conclusion}
We have studied the hard diffractive production of heavy vector mesons in deep
inelastic scattering. The hardness is provided by a large momentum transfer,
$|t|$, and allows one to probe the essential dynamics which determine the
pomeron of perturbative QCD. We have made predictions for the cross section
for a range of kinematical configurations accessible at HERA. Corrections to
the simple two-gluon picture of the pomeron have been computed using the
leading logarithmic formalism of BFKL and they are seen to be very large.

It is encouraging that the total cross section for this process is not too
small. Specifically, for $\surd s \approx 200$ GeV, photon
virtuality  $Q_{\gamma}^2 \approx 10$ GeV$^2$, $x' \gapproxeq 0.01$
and $|t| \geq 2$ GeV$^2$ (i.e. a mean $z$ of 0.6)
the total cross section $\sigma^T(\gamma^*p\to V+X)$ for the hard
diffractive production of $J/\Psi$ mesons is $\approx 1.5$ nb.  This is
to be compared with the asymptotic BFKL prediction of $\approx 3$ nb
and the Born prediction of $\approx 84$ pb. Although this cross
section (1.5 nb) is small, (it is only $0.01 \%$ of the total DIS cross
section) it is not prohibitively so.

Although we have concentrated on the general case of non-zero $Q_{\gamma}^2$,
all of our results could equally well apply to photoproduction (where the
production rate is much higher). We need not worry about Sudakov or VMD
corrections due to the largeness of $\bar{q}^2$. For example, for $Q_{\gamma}^2
= 0$, $\surd s = 200$ GeV and $|t| \geq 2$ GeV$^2$ one is probing a mean $z$ of
0.8 and the cross section for $J/\Psi$ production off transverse photons is
$\approx 15$ nb. At $\surd s = 100$ GeV this cross section falls to $\approx 5$
nb.

It is an important asset of this process that one does not need to observe the
products of the proton dissociation in order to extract the momentum transfer,
$t$. In DIS, observation of the scattered electron together with the decay
products of the vector meson allows a clean determination of $t$ whilst in
photoproduction one can assume that the incoming photon is collinear with the
incoming electron and hence the $p_T$ of the vector meson gives $t$.
Consequently, the rapidity gap between the vector meson and the scattered
parton can be as big as 9 units. As a result, HERA can probe the region where
the BFKL series is poorly convergent, i.e. where the resummation of leading
$\ln s$ terms is most important.

We restricted ourselves to heavy mesons for reasons of simplicity although
$\rho$ (or $\phi$) production will also provide important information on the
QCD pomeron with generally a
much higher production rate. Our calculations have been
totally general and so one could use them to predict the production
cross sections for any vector meson.
However, we expect the non-relativistic approximation to be
less appropriate for the light mesons. In addition, VMD contributions may
well start to become significant in the case of photoproduction for
not-too-large $t$. It is interesting to note that for
$|t| \gg m_V^2$ the yield of heavy vector mesons is larger than that of the
lighter ones. This is because the cross section for scattering transversely
polarised photons is proportional to $\Gamma^V_{e^+e^-}m_V^3$
(see eqs.(1),(3)). Since the electromagnetic widths of the $J/\Psi$ and $\rho$
are similar we therefore find the $J/\Psi$ rate to be enhanced by a factor
$\sim 50$ over the $\rho$ production rate. This is a simple consequence of the
fact that at large $|t|$, the pomeron has
enhanced coupling to the more compact $c\bar{c}$ pair which
constitutes the $J/\Psi$. However, this enhancement is diluted to a single
power of $m_V$ in the case of longitudinal photon scattering, which dominates
at large enough photon virtualities. We note that changing the photon
virtuality allows one to scan through the vector meson wave function (as
was discussed in ref.\cite{kz}).

Corrections to our predictions which arise from higher order QCD and
non-perturbative effects (in the dissociation of the proton) have been
estimated and should not be essential over the HERA range, perhaps
contributing by no more than $\approx 5\%$. However, higher order corrections
to the BFKL amplitude itself may be more significant. For example, if we take
$y = \ln [\hat{s}/(4 {\mathrm{Max}}(Q_{\gamma}^2,m_V^2,Q^2))]$ (rather than $y
= \ln (\hat{s}/4\bar{q}^2)$) then the cross sections quoted in these
conclusions are reduced by $\sim 40\%$. This ambiguity in the definition of $y$
is a direct consequence of the leading log nature of the calculation.

It should not be long before the HERA experiments are able to study data in the
regime considered in this paper. In particular, there is already some
preliminary data on both $J/\Psi$ and $\rho$ production out to $|t|$ values as
large as 1.5 GeV$^2$ (in DIS and photoproduction) and at $\gamma p$
centre-of-mass energies of $\approx 100$ GeV \cite{HERA}.

\vspace{10mm}

\noindent{\bf Acknowledgements:}  We should like to thank Prof. M.Arneodo,
Dr. H.Jung, Prof. C.Peroni and Dr. P.J.Sutton for many fruitful discussions.
Very special thanks to Prof. J.Bartels, H.Lotter and Dr. M.W\"usthoff for their
interest in this work. One of us (M.R.) gratefully acknowledges the
hospitality of the Torino University Physics Department, the Theory Group of
the RAL Particle Physics Department and the DESY
Theory Division, where part of this work was done.  

\newpage

\vskip 1.0cm
\begin{flushleft}
{\large\bf Figure Captions}
\end{flushleft}

\begin{description}
\item[Fig.\ 1] Feynman graphs corresponding to diffractive vector meson
production in DIS.

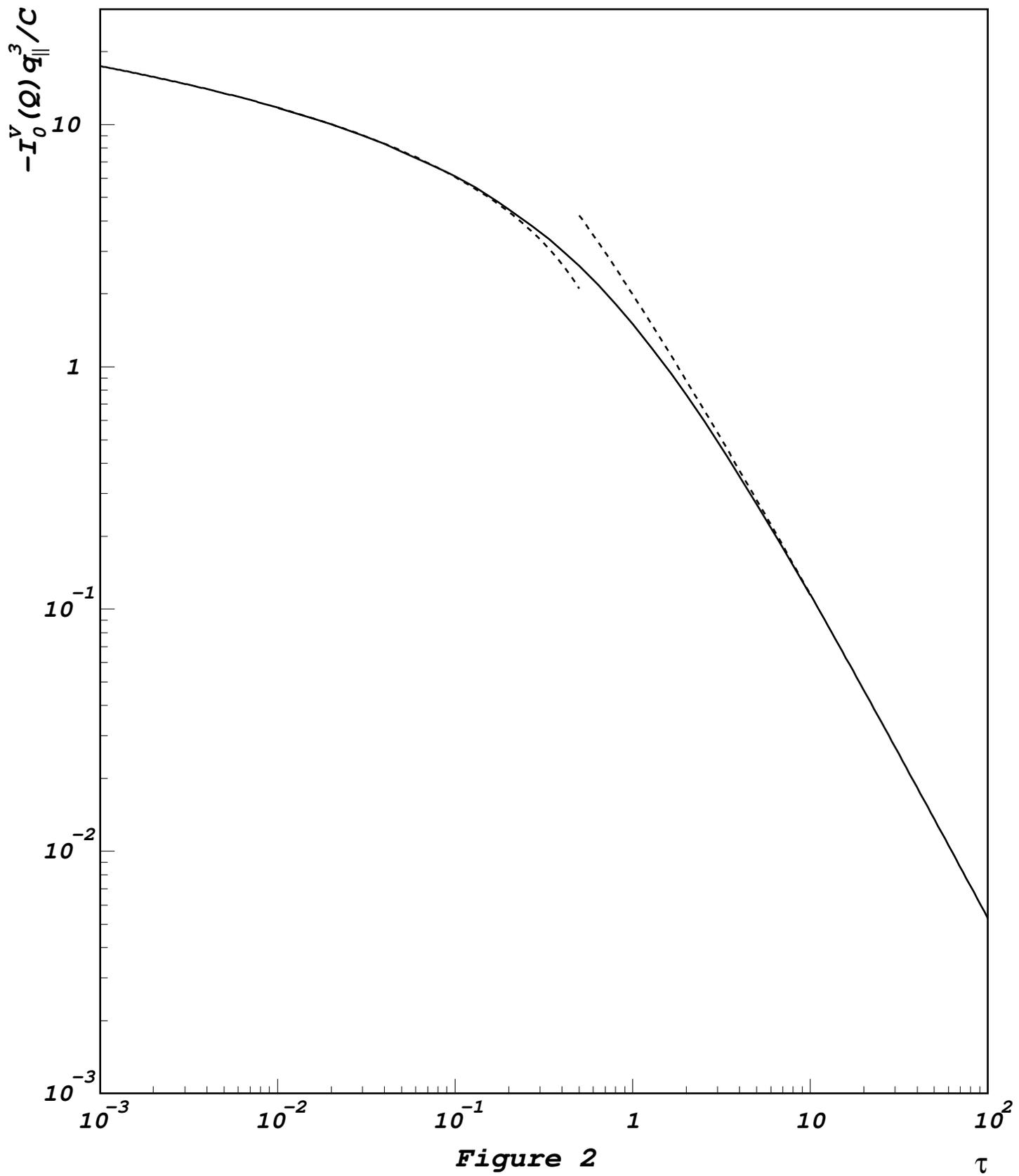
\item[Fig.\ 2] The asymptotic prediction for the diffractive scattering
amplitude. The dotted lines are the analytical approximations and the solid
line is the result of exact numerical computation.

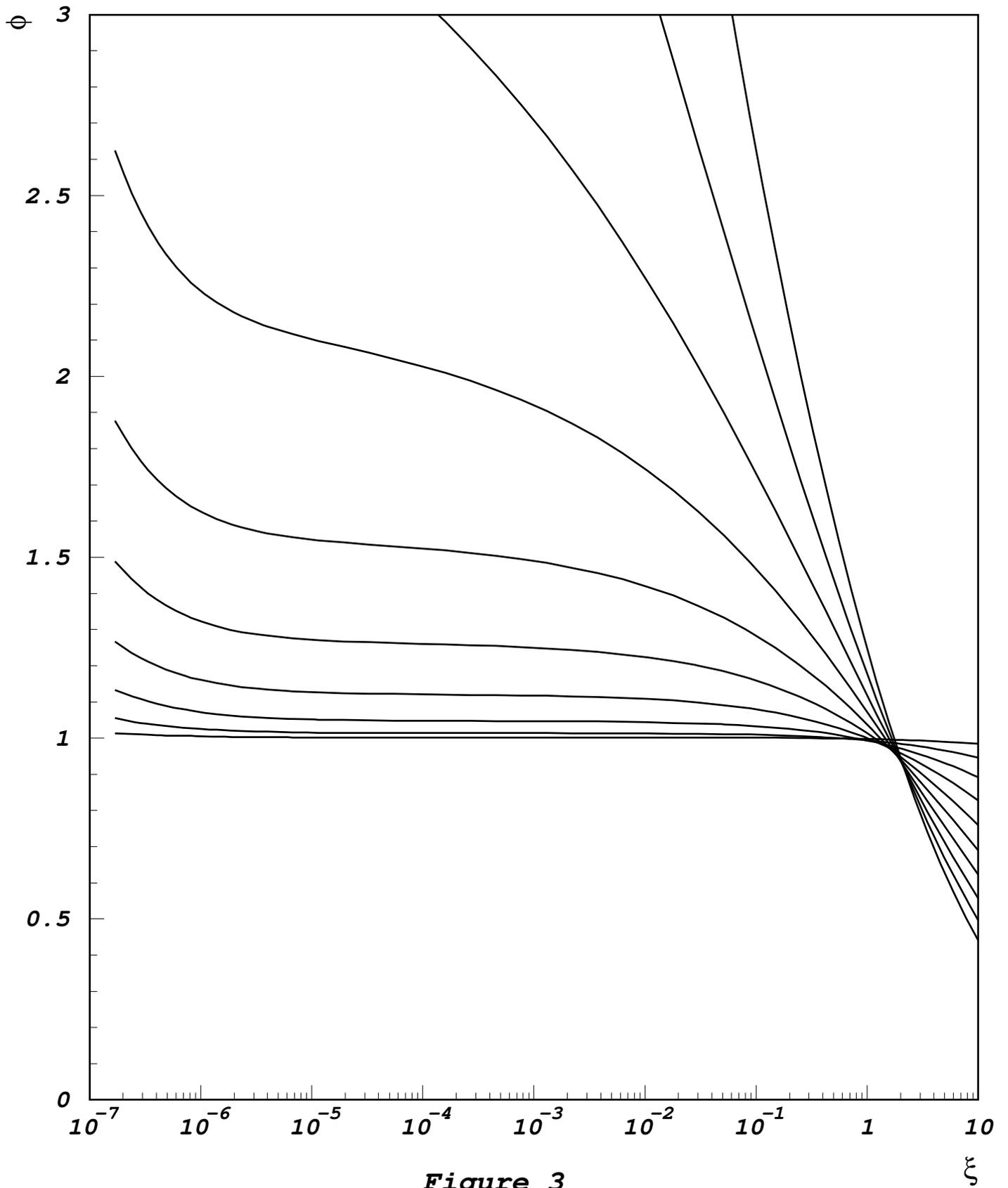
\item[Fig.\ 3] The deviation of solution of the BFKL equation in the low
$\xi$ region from the double logarithmic form (corresponding to
$\phi = 1$). The curves correspond to different $z$ values. At $\xi
=10$, the curves increase with decreasing $z$ from $z=1$ to $z=0.1$ in
steps of $0.1$.

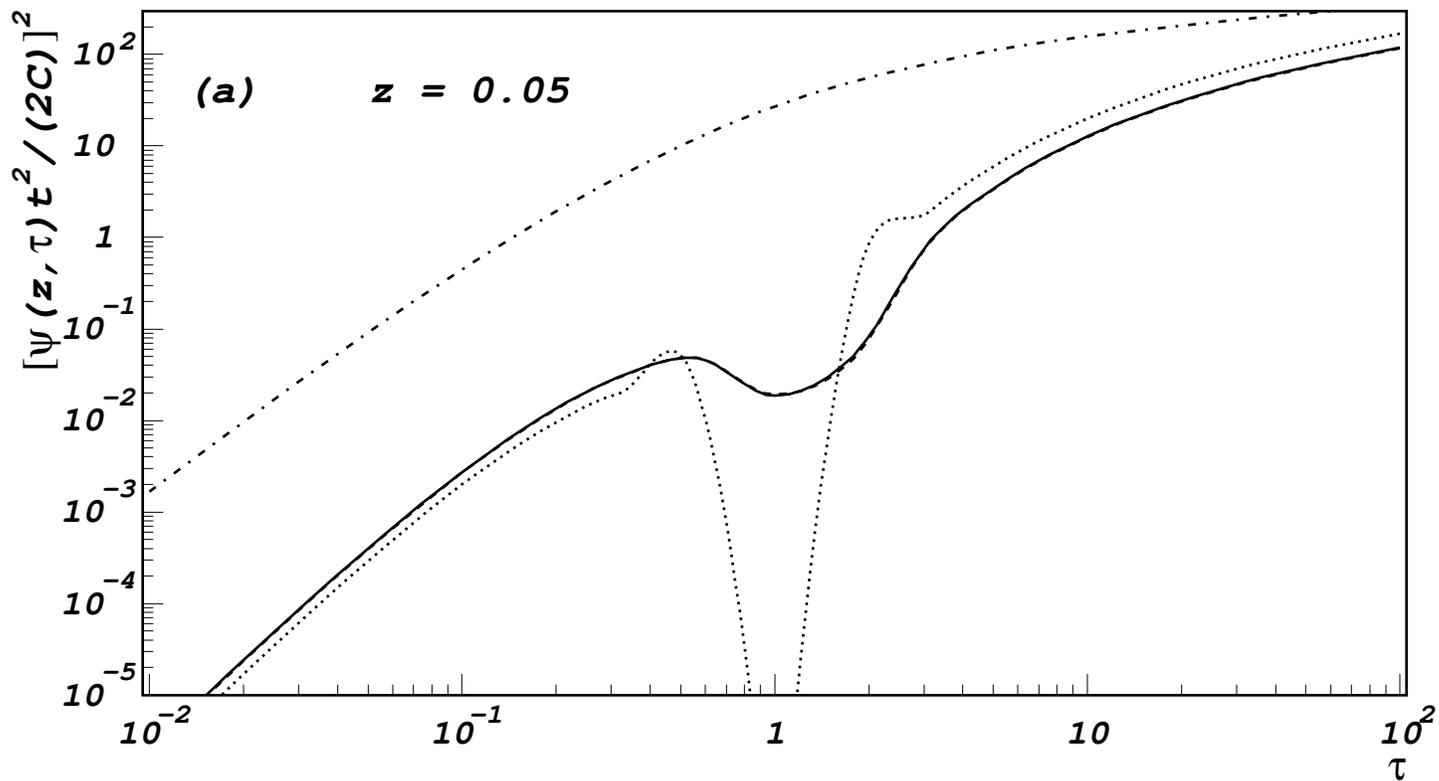
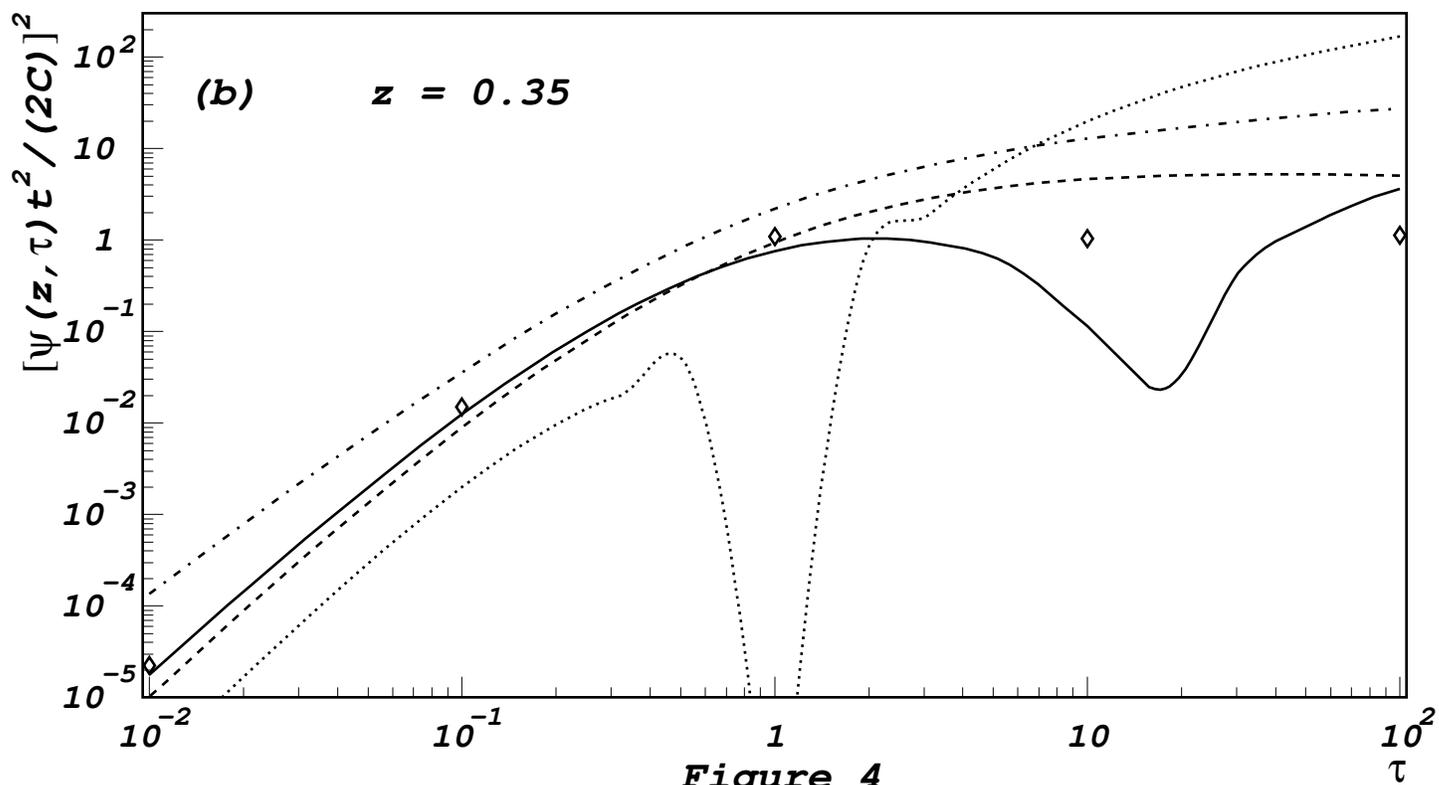
\item[Fig.\ 4] The peculiarly normalised cross section for $\gamma^* q \to Vq$,
calculated numerically by successively iterating the BFKL kernel. The dotted
line corresponds to two gluon exchange, the dashed line incorporates the
corrections predicted by the first iteration (of the BFKL kernel), the solid
line is the prediction after the second iteration and the diamonds are the
predictions after the third iteration. The dash-dot line is the asymptotic
result obtained by computing eq.(9) numerically. No resummation of the double
logarithms has been performed. (a)$z = 0.05$; (b) $z=0.35$.

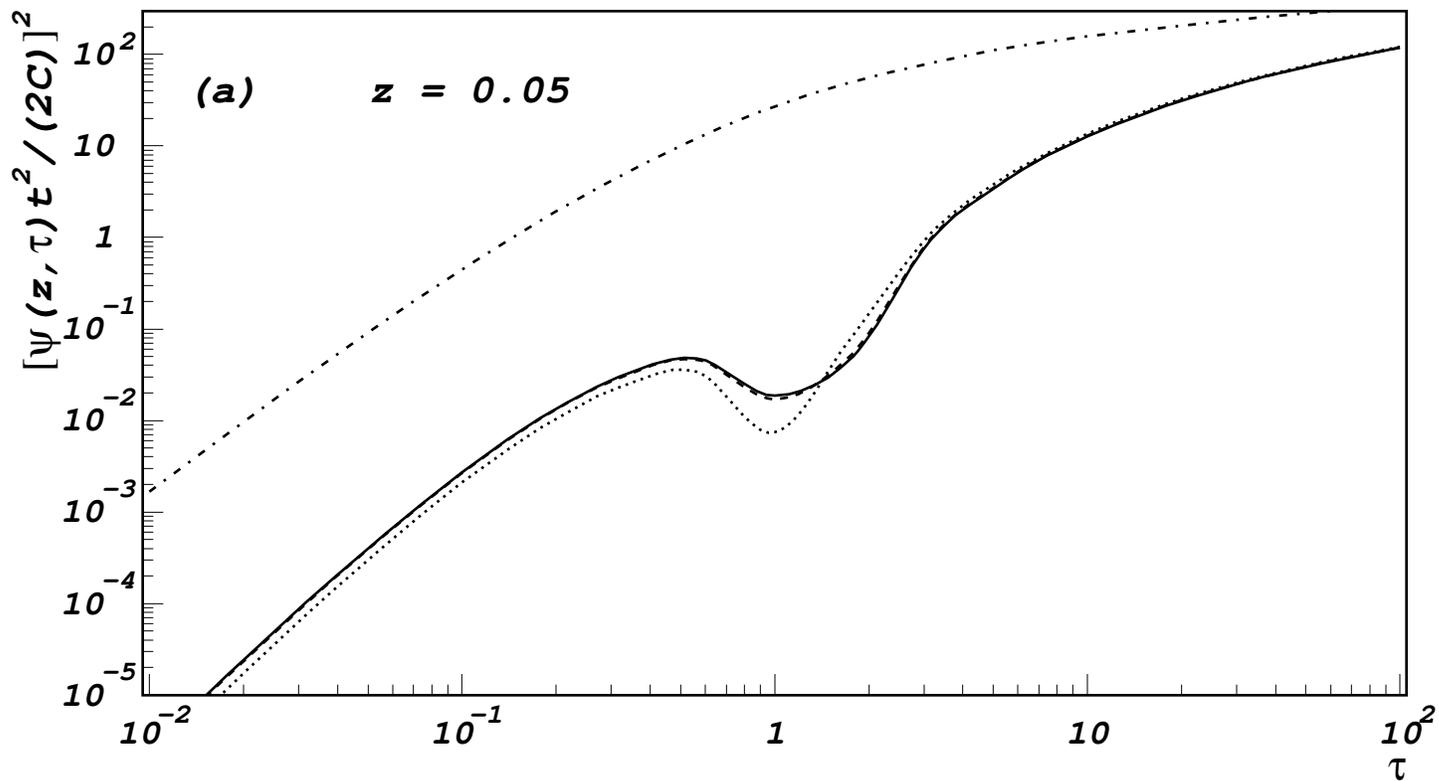
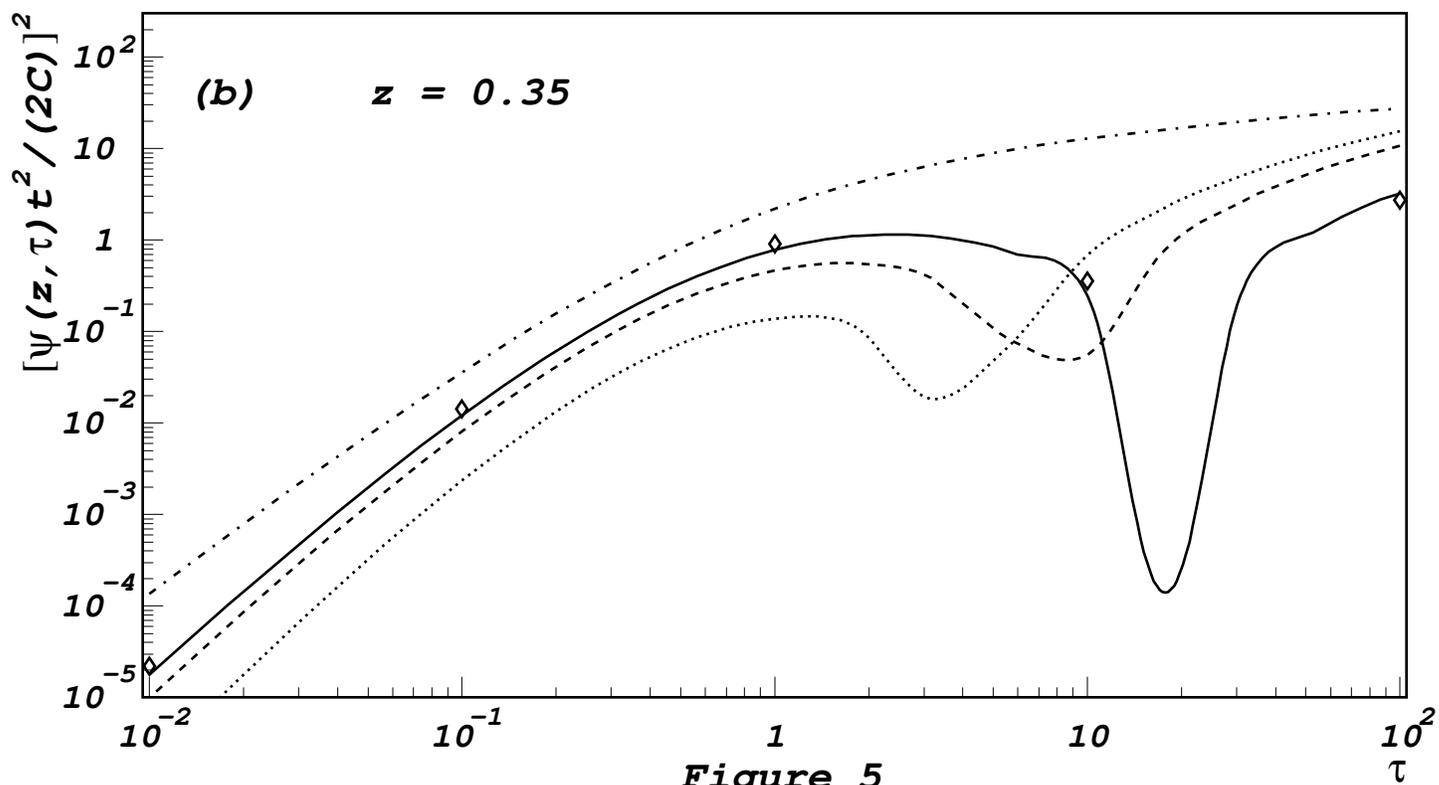
\item[Fig.\ 5] As in fig.4 except that the double logarithms have been
resummed in the manner described in the text.

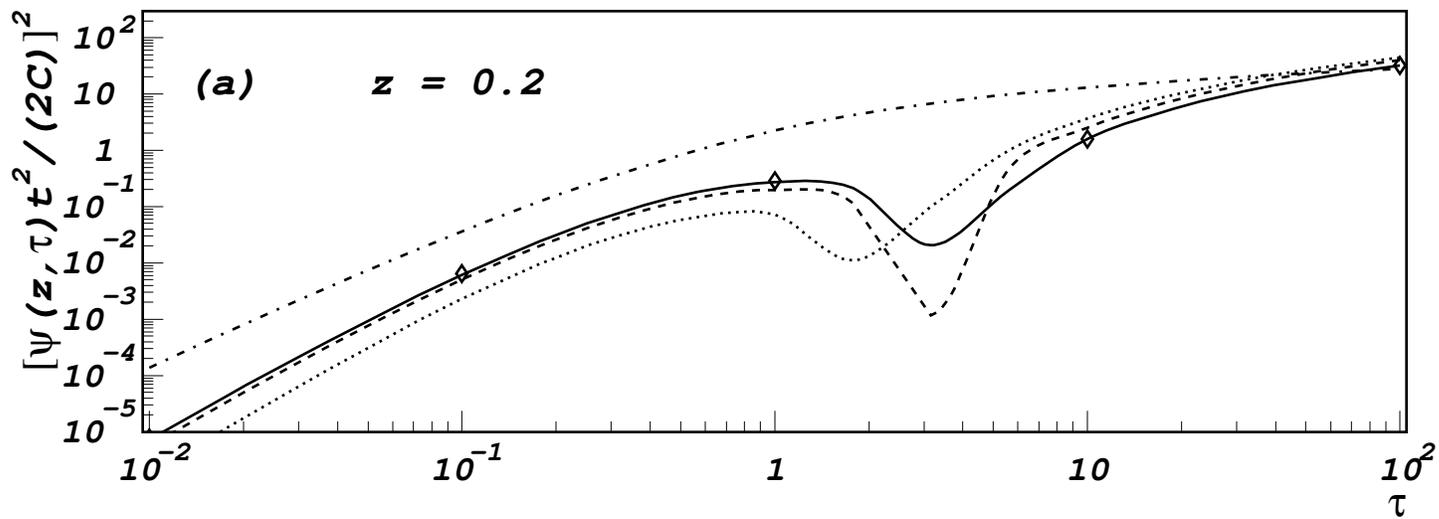
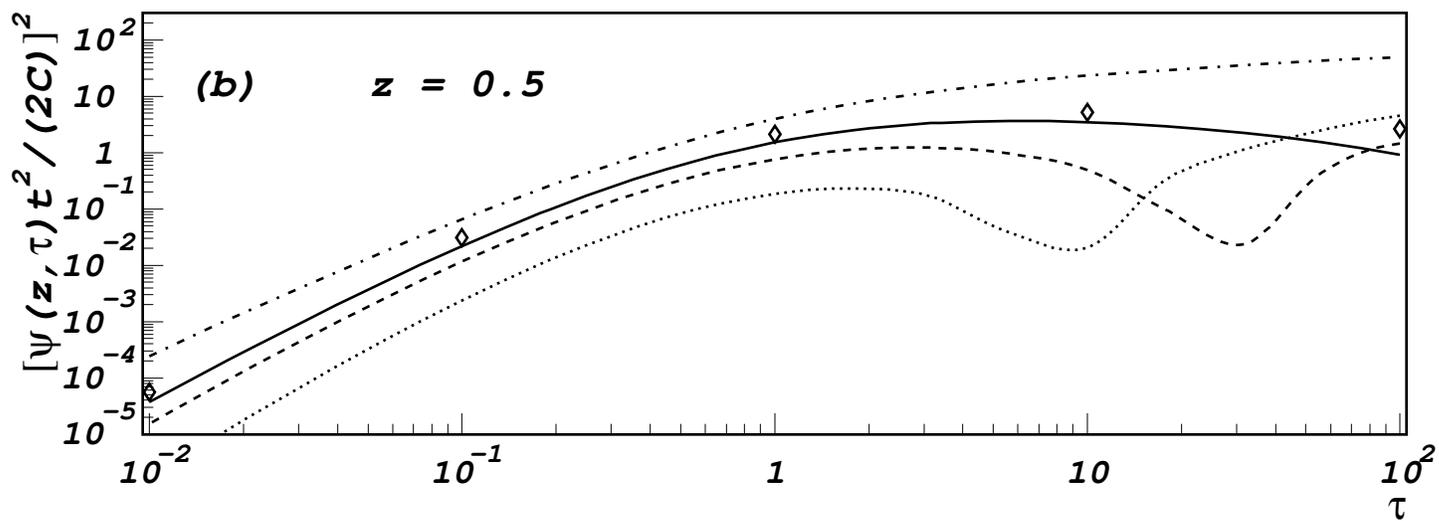
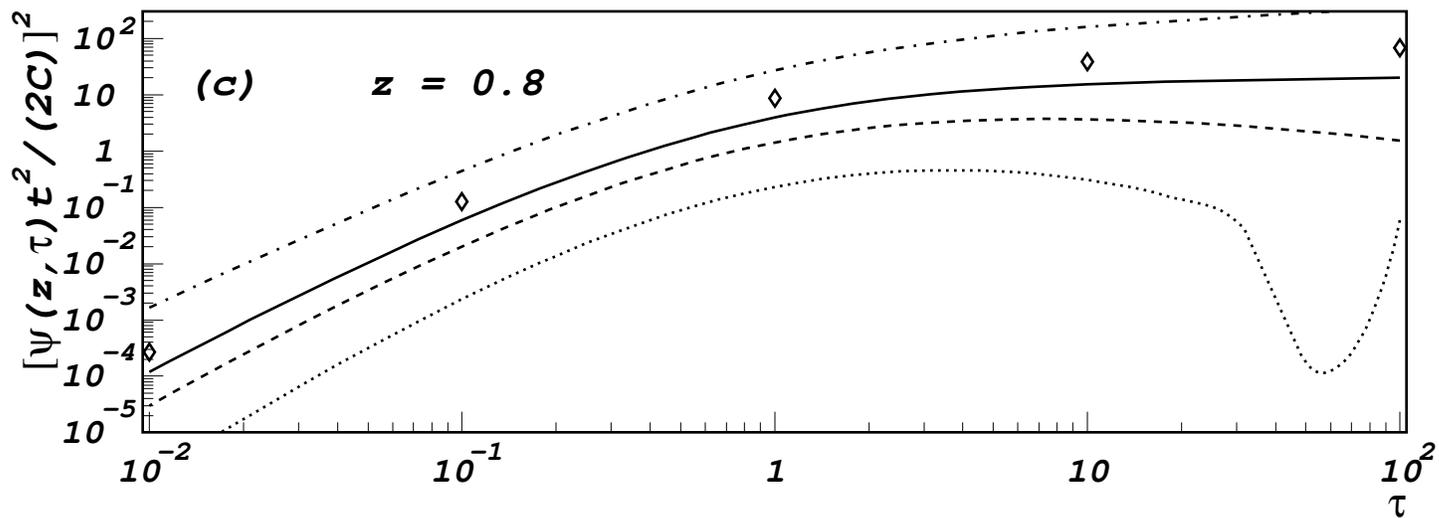
\item[Fig.\ 6] As in fig.5, but for (a) $z=0.2$; (b) $z=0.5$; (c) $z=0.8$.

\end{description}


\end{document}